\documentclass[a4paper,11pt]{article}
\pdfoutput=1 % if your are submitting a pdflatex (i.e. if you have
\usepackage{jcappub}
\usepackage[mathscr]{euscript}
\usepackage{upgreek}
\usepackage{slashed}
\usepackage{cancel}

\title{\LARGE{Exorcising the Ghost Condensate Dark Energy with a Sextic Dispersion Relation}}

\author{A. Ashoorioon} \author{and A. Yousefi-Sostani  }
 \emailAdd{amjad@ipm.ir} \emailAdd{yousefiae@ipm.ir}
\affiliation{School of Physics, The Institute for Research in Fundamental Sciences (IPM)\\P.O. Box 19395-5531, Tehran, Iran}

\abstract{The universe's current acceleration is a pretty recent phenomenon in cosmological time scales. This means that the modes that have left our horizon since the beginning of the contemporary acceleration phase, have not really reached the exact IR limit. Noting this observation, we reconsider the possibility of having a ghost condensate as dark energy with a sixth-order dispersion relation. Looking at the three-point function of such a theory, we obtain the constraints on the coefficient of the sixth-order dispersion relation to avoid strong coupling. Such a ghost condensate if coupled to the standard model fields, induces a constant Lorentz-violating spin-dependent force, which can gravitate or anti-gravitate.}

\subheader{IPM/P-2023/}
\begin{document}
\maketitle
\flushbottom
\section{Introduction}\label{sec1}

The discovery of the current acceleration of the universe took the physicist by surprise at the end of the last century.\cite{SupernovaSearchTeam:1998fmf} The simplest explanation for this phenomenon is considering the cosmological constant which has to be extremely small. Even from the anthropic perspective, the discovered value of the cosmological constant is two orders of magnitude below the anthropic upper bound \cite{Weinberg:1987dv}. Hence it is difficult to explain the origin of such a small scale. Another approach to solving this problem is the modification of gravity at large scales. Among these approaches, one can consider varieties of scalar-tensor theories \cite{Brans:1961sx,Dicke:1961gz,Clayton:1998hv} which only modify the long-range interaction between two masses by a constant force. There are other ways one can modify gravity in which a scalar degree of freedom, $\pi$, interacts with the matter with gravitational strength, and at scales smaller than the length scale of Infrared (IR) modification, the theory reduces to a scalar-tensor theory rather than general relativity. One of these theories is DGP braneworld model \cite{Dvali:2000hr} and the other one is the Fierz-Pauli theory of massive gravity \cite{Fierz:1939ix}. In these theories,  the additional scalar degree of freedom $\pi$ only becomes healthy and obtains a kinetic term due to coupling with gravity. The price one pays is that the theory becomes strongly coupled at Ultraviolet (UV) cutoff $\Lambda_{\mathrm{UV}}$. 

Contrary to the above classes of modifications of gravity, \cite{arkani01,Arkani-Hamed:2003juy} suggested a ``ghost condensate" theory in which there is no strong coupling for the scalar $\pi$ at $\Lambda_{\mathrm{UV}}$, even in the absence of coupling to the gravity. The theory resembles the cosmological constant, but unlike that, it has physical excitations around a spontaneously Lorentz-breaking background $\langle \dot{\phi}\rangle=M^2 $ that breaks the time-translation  invariance too. This is like the Higgs mechanism  in gauge theories, but here instead of getting massive gravitons with five polarizations, due to the Lorentz-breaking nature of the theory, one would get 
only one additional scalar field with modified dispersion relation. Unlike the gauge theories, where endowing them with mass, one would get an exponential Yukawa-like suppression, here one would get an oscillatory modulation of the potential. The kinetic term for the ghost field appears with a sign opposite to the standard model fields \footnote{Spacetime metric signature is mostly-negative $(+ \ - \ - \ -)$.}. 
\begin{equation}\label{s01eq01}
\mathcal{L}=-\frac{1}{2}\partial^{\mu}\phi\partial_{\mu} \phi+\cdots\,.
\end{equation}
Like how in the Higgs mechanism the quartic correction stabilizes the quantum instability in $\phi=0$, we can envisage that higher order terms like $(\partial\phi)^4$ stabilize the theory. Of course, the analogy is not complete, since the $(\partial \phi)^4$ term is nonrenormalizable. Still, such a description is viable for a systematic low-energy expansion of the fluctuations around the ghost background. 

One point that \cite{arkani01} emphasized is that going beyond the quartic dispersion relation, $\omega^2\propto k^4$ causes strong coupling in the IR limit. The argument goes as this: if one assumes that $E\rightarrow sE$ and consider for example a dispersion relation like $\omega^2\propto k^6$, the scaling dimension of $t$, $x$ and $\pi$ would be as follow:
\begin{eqnarray*}
  t &\rightarrow & s^{-1} t \\
  x &\rightarrow & s^{-1/3} x \\
  \pi &\rightarrow & s^{0} \pi
\end{eqnarray*}
Therefore the interacting operators 
\begin{equation}\label{s01eq02}
\int \mathrm{d}^3 x~\mathrm{d} t~M^4 \dot{\pi} (\nabla\pi)^2\,,
\end{equation}
scales like $s^{-1/3}$. This means that the interacting operator becomes relevant, and the theory becomes strongly coupled at low energies. This certainly invalidates the Lorentz-violating EFT with $\omega^2\propto k^6$ for describing the inflation as during inflation there are modes that approach the superhorizon limit in the mathematical sense, $k/aH\rightarrow 0$.\footnote{$a(t)$ stands for the scale factor of the FLRW background and $H= \dot{a}/a$ is called the Hubble parameter, $H_0$ stands for today's value of the Hubble parameter. Also, the dot $\dot{}$ denotes the derivative \textit{w.r.t.} the physical time $t$.} Even during inflation if before the EFT before the mode with $\omega^2\propto k^6$ makes a transition to $\omega^2\propto k^4$ or $\omega^2\propto k^2$, before it makes becomes strongly coupled at the scale $\Lambda^{\mathrm{IR}}_6$, one can have a consistent EFT for describing inflation \cite{ashoorioon18,Ashoorioon:2018ocr,Ashoorioon:2021srt}. The question now arises is that what happens if such an EFT is to describe the late time acceleration of the universe. Late-time acceleration of the universe is a recent event in today's cosmological time scales, so it might be that one can describe the current acceleration of the universe with a ghost dark energy with pure $\omega^2\propto k^6$. In this work, we entertain this idea and try to see to what extent our theory is fine-tuned. We find out that such a theory can describe late-time acceleration without that much fine-tuning.

We also find out that if such a ghost condensate, with $\omega^2\propto k^6$ dispersion relation, has direct coupling to the standard model fields through a derivative coupling of the ghost to the axial vector currents, a constant spin-dependent force mediated by the Nambu-Goldstone boson exchange is induced. On the other hand, if the ghost sector does not have any direct coupling to the standard model fields and only couples gravitationally to them, the length and time scale it takes for a gravitational phenomenon to propagate are
\begin{equation}
    r_c\sim \frac{M_\text{Pl}}{M^2}\,, \qquad t_c\sim \frac{M_\text{Pl}^3}{M^4}\,.
\end{equation}
For example, if a cloud of dust collapses to form a star, the modification of gravity takes place at scales $r_c$, only after the time $t_c$. Please note that for a ghost sector with $\omega^2\propto k^4$, 
although the length scale  $r_c$ is given by $\frac{M_{\mathrm{ Pl}}}{M^2}$, the time scale for the change to happen is given by $t_c\sim \frac{M_{\mathrm{ Pl}}^3}{M^4}$. However, like the case with $\omega^2\propto k^4$, the modification to Newton's potential is oscillatory at distances of order $r_c$. For the ghost condensate to describe the contemporary acceleration $M\sim 10^{-3}$ eV. Although this means that modification of gravity happens at $r_c\sim H_0^{-1}$, the time scale for seeing the oscillatory modification is $t_c\sim 10^{70}$ years. So no cosmological experiment can tell apart the ghost condensate dark energy and the cosmological constant. 

The structure of the paper is as follows: First, we spell out the structure of the ghost condensate with a sextic dispersion relation. We study the ghost condensate with the sixth-order dispersion relation in the absence of gravity. Then we show that the theory makes sense to describe the late-time acceleration of the universe if some constraints are assumed on the coefficient of the dispersion relation. Then we discuss the possibility of the interaction of the ghost condensate with the sixth-order dispersion relation with the standard model fields and show that this will lead to spin-dependent but distance-independent force  between the particle. Finally, we discuss the possibility of such a  ghost condensate coupled to gravity. 

%%%%%%%%%%%%%%%%%%%%%%%%%%%%%%%%%%%%%%%%%%%%
\section{Ghost Condensate with the Sextic Dispersion Relation} \label{sec2}

Let us consider the following action, which is dependent on $\phi$ only through its derivatives,
\begin{equation}\label{s02eq01}
     S  = {S}_{0} + \Delta {S}\,,
\end{equation}
where
\begin{align} 
    {S}_{0} =\int \mathrm{d}^4 x  \sqrt{-g} & \left[M^4 P(X) \right],\label{s02eq02} \\
    \Delta {S} =\int \mathrm{d}^4 x \sqrt{-g} \left[ M^2 \left(S_{1} (\Box \phi)^2  + S_{2} (\partial_\mu \partial_\nu \phi)^2 \right) + \right. & \left. S_{3}\left(\partial_{\mu} \square \phi\right)^{2}+S_{4}\left(\partial_\mu \partial_{\nu} \partial_{\rho} \phi\right)^{2} + \cdots \right] \label{s02eq03}. 
\end{align}
$g_{\mu \nu}$ denotes the spacetime metric with the determinant $g$. The coefficients $S_{i}\,,i=1 \cdots 4$ are also dependent only on $X = \partial_{\mu}\phi\partial^{\mu}\phi$. The above action is written assuming $\phi \rightarrow -\phi$ symmetry for simplicity. This prevents terms with odd numbers of $\phi$ in the action. 

The action $S_0$ provides terms, including the ghost part of the theory. It contains the higher order terms like  $(\partial\phi)^4$ that stabilize the theory. Such terms are non-renormalizable, and it seems that the description of the ghost condensate in terms of an effective field theory is not possible in this way, but it can be shown that one can still have a meaningful theory of fluctuations, $\pi$, around the ghost condensate background. As we will see, the  additional terms proportional to $S_i$ provide higher-order derivatives of $\pi$.

For now, varying just the action \eqref{s02eq02}, one finds the equation of motion (EOM)
\begin{equation} \label{s02eq04}
\partial^\mu \left[P'(X) \partial_\mu  \phi\right] =0\,, 
\end{equation}
which yields a solution $\partial_{\mu}\phi=\mathrm{constant}$, when gravity is ignored. A time-like Lorentz-violating solution to the above EOM is $\phi = ct$ where $c$ is an arbitrary dimensionless constant. Such a solution breaks the $\phi$ shift symmetry and time translation symmetry to an unbroken diagonal shift symmetry. Let us look at small fluctuations around this solution,
\begin{equation} \label{s02eq05}
\phi = ct + \pi,
\end{equation}
where $\pi$ is the Nambu-Goldstone mode of the broken symmetry.
The second (and higher) order action for the fluctuation around this solution is
\begin{equation} \label{s02eq06}
\begin{split}
S_0 &=  \ M^4 \int \mathrm{d}^4 x \sqrt{-g} \Big[ \dot{\pi}^2 ( P'(c^2) + 2 c^2 P''(c^2) ) - P'(c^2) (\partial_i \pi)^2 - 2 P''(c^2) c \dot{\pi} (\partial_i \pi)^2   \\ &+ \frac{1}{2}P''(c^2)(\partial_i \pi)^4  + \cdots \Big]\,.
\end{split}
\end{equation}
The time and spatial kinetic sign of the fluctuations $\pi$ appear with the standard sign if $c$ is such that
\begin{equation} \label{s02eq07}
P'(c^2) > 0, \text{\hspace{0.5cm}} P'(c^2) + 2 c^2 P''(c^2)  > 0.
\end{equation}

Any value of $c$ that satisfies the relation \eqref{s02eq07} is allowed. Nevertheless, it can be shown that in an expanding background that $P'(\dot{\phi}^2)\propto a(t)^{-3}$. This means that for the special value $c_{*}$, where $\dot{\phi}^2 = c_{*}^2$, the coefficient of the spatial quadratic term, $(\partial_i \pi)^2$ vanishes.

Considering the additional contributions to the action, \eqref{s02eq03} higher order spatial derivative terms for $\pi$ appear. When one considers just the terms $S_{1,2}$, the low-energy dispersion relation of $\pi$ will be given as $\omega^2 \sim k^4$ in the absence of gravity. This is the original  ghost condensate with quartic dispersion relation proposed in \cite{arkani01}. However, we are intended to find a sextic low-energy dispersion relation. So we have added the terms proportional to $S_{3,4}$. One can expand the action around $c_*^2$ to obtain such an effective low-energy action. If we put $P''(c_*^2) = \frac{1}{4c_*^2}$ by rescaling $P$ and $M^4$, then the action \eqref{s02eq01} is obtained
\begin{equation}\label{s02eq08}
S = \int \mathrm{d}^4 x  \sqrt{-g} \left[ \frac{1}{2}M^4\dot{\pi}^2 - \frac{1}{2 }\Bar{M}^2 (\partial_i \partial_j \pi)^2  -\frac{1}{2 }\bar{S}(\partial_i \partial_j \partial_k \pi)^2 + \cdots \right]\,,
\end{equation}
where $\Bar{M}^2 = -2 M^2 (S_{1}(c_*^2) + S_{2}(c_*^2))$ and $\bar{S}= - 2 (S_{3}(c_*^2) + S_{4}(c_*^2))$. It is assumed that $S_{1}(c_*^2) + S_{2}(c_*^2)<0$ and $S_{3}(c_*^2) + S_{4}(c_*^2)<0$ to stabilize the fluctuations around $\phi=c_{*}t$ . The canonical $\pi$ mode will obtain the following low-energy dispersion relation
\begin{equation} \label{s02eq09}
    \omega^2=\frac{\bar{M}^2}{M^4} k^4+\frac{\Bar{S}}{M^4} k^6\,.
\end{equation}
In expanding spacetime, the action \eqref{s02eq08} takes the form
\begin{equation}\label{s02eq10}
S= \int a^3  \left[ \frac{1}{2}M^4\dot{\pi}^2 - \frac{1}{2}\Bar{M}^2 \left(\frac{\nabla^2}{a^2} \pi\right)^2  -\frac{1}{2}\bar{S}\left(\frac{\nabla^3}{a^3} \pi\right)^2 + \cdots \right]
\end{equation}
The low-energy dispersion relation in expanding background will be dominant by the $k^6$ dispersion relation if
\begin{equation}\label{s02eq12}
    \bar{S} \frac{k^6}{a^6} \gg \bar{M}^2 \frac{k^4}{a^4} \implies \frac{k^2}{a^2} \gg \frac{\Bar{M}^2}{\bar{S}}
\end{equation}
Noting that the onset of domination of dark energy is $z_c\simeq 0.3$,\cite{Weinberg:2013agg} the minimum value of the left-hand side of the above inequality is $\frac{H^2}{1.69}$ and thus
\begin{equation}\label{s02eq13}
    \bar{S} \gg \frac{\Bar{M}^2}{H^2 (z + 1)^2} \implies \bar{S} \gg \frac{\Bar{M}^2}{1.69 H^2}.
\end{equation}
Under this condition, the action \eqref{s02eq09} effectively becomes
\begin{equation}\label{s02eq14}
S\simeq \int a^3  \left[ \frac{1}{2}M^4\dot{\pi}^2  -\frac{1}{2}\bar{S}\left(\frac{\nabla^3}{a^3} \pi\right)^2 + \cdots \right],
\end{equation}
and the corresponding dispersion relation is
\begin{equation}\label{s02eq15}
    \omega^2 \simeq  \frac{\bar{S}}{M^4}\frac{k^6}{a^6}.
\end{equation}
As we mentioned in section \ref{sec1}, the energy scaling implies that the above effective field theory breaks down when $k \rightarrow 0$. However, applying the theory to the current acceleration of the universe, the smallest $k$ that has exited the horizon in the latest acceleration phase is $\simeq \frac{H}{1.3}$. We will investigate under what conditions the strong coupling would have not occurred for such a theory.

\section{Viability of \texorpdfstring{$\omega^2 \sim k^6$}{} Dispersion Relation}

In this section, we compute the three-point function for the Nambu-Goldstone boson, $\pi$, and demand that it does not lead to strong coupling for the modes that have exited the horizon during the recent acceleration.  The smallest $k$-mode that has exited the horizon currently has the wavenumber $k\simeq H/(1+z_c)$, where $z_c$ is the redshift of domination of the dark energy. Like inflation, one can quantify the avoidance of strong coupling as 
\begin{equation}\label{s03eq01}
\left|f_\text{NL}\right| \ll |\zeta|^{-1},
\end{equation}
where $\zeta=-H\pi$, and the $f_\text{NL}$ is defined as follows
\begin{equation}\label{s03eq02}
f_\text{NL} \sim \frac{B(k_1,k_2,k_3)}{\Delta_{\zeta}^4(k)}\,.
\end{equation}
Above, 
\begin{equation} \label{s03eq03}
\Delta^2_{\zeta}(k)=\frac{k^3}{2\pi^2}\langle\zeta\zeta\rangle
\end{equation}
First, we compute the two-point function of the curvature perturbation  $\zeta = - H \pi$. For this purpose, we introduce the conformal time $\eta = \int {\mathrm{d}t}/{a(t)} $ and the new variable $u(\eta) = a(\eta) \pi_c(\eta )$ defined in terms of the canonical variable $\pi_c / M^2 = \pi $.  During the late-time accelerating phase of the universe, the background could be approximated with the de Sitter metric,
\begin{equation}\label{s03eq04}
\mathrm{d}s^2 = a^2(\eta) (\mathrm{d}\eta^2 - \delta_{ij} \mathrm{d}x^i \mathrm{d}x^j) 
\end{equation}
where $a(\eta)=-1/(H \eta)$. We assume that today $a(\eta_{0})=1$, therefore $\eta_0=-1/H$. With these considerations, we obtain the equation of motion for $u_k(\eta)$, the Fourier transform of the variable $u(\eta)$ to be
\begin{equation}\label{s03eq05}
\left(-\frac{a''(\eta )}{a(\eta )}+\frac{\bar{S}}{M^4} \frac{k^6}{a(\eta )^4}\right) u(\eta ) +u''(\eta )=0.
\end{equation}
which in the de-Sitter spacetime has the solution,
\begin{equation}\label{s03eq06}
u(\eta , k) =a\pi= -\sqrt{\frac{3}{2}} \frac{M}{\bar{S}^{1/4} k^{3/2}H \eta} \exp \left(\frac{-i \sqrt{\bar{S}}}{3}\left(\frac{H}{M}\right)^2(k \eta)^3 \right).
\end{equation}
The dimensionless two-point function is obtained to be
\begin{equation}\label{s03eq07}
\Delta^2_\zeta (k) = \frac{3 \pi }{ \sqrt{\bar{S}}} \left(\frac{H}{M}\right)^2 .
\end{equation}

We now exploit the in-in formalism to compute the three-point function \cite{Maldacena:2002vr}
\begin{equation}\label{s03eq08}
\begin{split}
 \langle \zeta ( {k}_1) \zeta( k_2) \zeta( k_3)\rangle  &  = -\frac{H^3}{M^6} \text{Re} \left\lbrace -2i \langle \pi_c (\tilde{\eta}, \vec{k}_1) \pi_c(\tilde{\eta}, \vec{k}_2) \pi_c(\tilde{\eta}, \vec{k}_3) \int \frac{\mathrm{d}^3 q_1}{(2 \pi)^3}\frac{\mathrm{d}^3 q_2}{(2 \pi)^3}\frac{\mathrm{d}^3 q_3}{(2 \pi)^3} \right. \\ & \left. \times \int \mathrm{d}^3 x \int^{\eta_0}_{\eta_1} \mathrm{d}\eta' \ a(\eta')^4 \ \mathcal{H}_\text{int} e^{-i(\vec{q}_1 + \vec{q}_2 + \vec{q}_3) \cdot \vec{x} } \rangle  \right\rbrace,
\end{split}
\end{equation}
where the corresponding interaction term with the most negative scaling dimension could be read off from \eqref{s02eq06}. It is $\mathcal{H}_\text{int}=-\mathcal{L}_\text{int} = \frac{1}{2c_* M^2}\dot{\pi}_c\left({\partial_i \pi_c}\right)^2$. The era of dominance of the late dark energy occurs around $z_c \sim 0.3$ \cite{Weinberg:2013agg}. As before, we can approximate this late-time acceleration with a quasi-de Sitter phase, and thus the integration upper and lower bounds will be respectively $\eta_0 = -\frac{1}{H}$ and $\eta_1 = -\frac{1 + z_c}{H}$. Using the solution \eqref{s03eq06}, one can do the integrations explicitly. One obtains 
\begin{equation}\label{s03eq09}
\begin{split}
 \langle \zeta (k_1) \zeta(k_2) \zeta(k_3)\rangle &= \\ & \frac{3^{8/3}(2 \pi)^3\delta\left(\Sigma_{i=1}^3 \vec{k}_i\right)}{16 c_{*} \bar{S}^{4/3}}\left(\frac{H}{M}\right)^{8/3} \\ & \times \text{Re} \left\lbrace \Gamma \left(\frac{2}{3},\frac{i \sqrt{\bar{S} }}{3}\left(\frac{H}{M} \right)^{2} \left( \Sigma_{i=1}^3  k^3_i \right) \eta^3 \right) ^{\frac{-1}{H}}_{\frac{-(1 + z_c)}{H}} \right\rbrace \\ &  \times \left[\frac{k_1^3(k_1^2 - k_2^2 - k_3^2)}{(k_1^3 + k_2^3 + k_3^3)^{2/3}}\frac{1}{(k_1 k_2 k_3)^3} +\text{perm.} \right]\,.
\end{split}
\end{equation}
Defining two new parameters, $\mathsf{K}^3 \equiv \sum_{i=1}^3 k^3_i$ and $\mathscr{K} \equiv k_1 k_2 k_3$, we have
\begin{equation}\label{s03eq10}
\begin{split}
B(k_1,k_2,k_3) = \frac{\left(3\frac{H}{M} \right)^{8/3}}{4 (2\pi)^4 c_* \bar{S}^{4/3}} & \text{Re} \left\lbrace\Gamma \left(\frac{2}{3},\frac{i \sqrt{\bar{S}}}{3}\left(\frac{H}{M} \right)^{2}\mathsf{K}^3 \eta^3\right)^{\frac{-1}{H}}_{\frac{-(1 + z_c)}{H}}\right\rbrace\\ & \times \left(\frac{ k_1^3(k_1^2 - k_2^2 - k_3^2)}{\mathsf{K}^{2}\mathscr{K}} + \text{perm.} \right)
\end{split}
\end{equation}
If one chooses the configuration in which all the modes are equal to the smallest mode that has left the horizon since the beginning of the acceleration, $k_1=k_2=k_3=H/1.3$, and replaces in the \eqref{s03eq10}, the inequality 
\eqref{s03eq01} takes the form
\begin{equation} \label{s03eq11}
    \frac{3\left(\frac{H}{M} \right)^{-4/3}}{(2\pi)^6 c_* \bar{S}^{1/3}} \left| \text{Re} \left\lbrace\Gamma \left(\frac{2}{3},\frac{i \sqrt{\bar{S}}}{3 (1.3)^3}\left(\frac{H}{M} \right)^{2} H^3\eta^3\right)^{\frac{-1}{H}}_{\frac{-(1 + z_c)}{H}}\right\rbrace \right| \ll \frac{\bar{S}^{1/4}}{\sqrt{3 \pi}} \left( \frac{H}{M} \right)^{-1}.
\end{equation}
Noting that Hubble parameter today is $H\sim 10^{-33}$ eV and $M\sim 10^{-3}$ eV it is reasonable to assume $\frac{H}{M} \ll 1$. Then one can expand the left-hand side of the above inequality to obtain
\begin{equation} \label{s03eq12}
    c_* \bar{S}^{1/4} \gg \frac{9 \alpha}{\sqrt{(2 \pi)^{11}}}\frac{H}{M}.
\end{equation}
where $\alpha = 1 - \left(1 + z_c\right)^{-3} $. For the current acceleration of the universe, $z_c \sim 0.3$, $H\sim 10^{-33}$ eV and $M\sim 10^{-3}$ eV, we will obtain the loose bound
\begin{equation}\label{s03eq13}
    c_* \bar{S}^{1/4}\gg 10^{-34}\,,
\end{equation}
which can easily be satisfied for $c_*$ and $\bar{S}$ of order one. This shows that strong coupling can easily be avoided for the $\omega^2 \propto k^6$ dispersion relation for the ghost dark energy.

Let us also investigate when the unitary bound is saturated, \textit{i.e.} when the the three-point function becomes comparable with the two-point function squared. Assuming the dominance of dark energy, the matter energy density redshifts. Thus, the Hubble parameter will reach an asymptotic value of around 56 Km/s/Mpc in the future, which still has the same order of magnitude as the Hubble parameter today. One can show that the unitary bounds will break down in the far future when the redshift of the dark energy dominance is of order  $z_{c}^{*} \geq 6 \times 10^{33}$, which is about 77 e-folds from now. 

\section{Coupling of the Ghost Field to the Standard Model Particles}

Although the standard model fields can avoid coupling to the ghost field, it would be interesting if such coupling exists. The story for the ghost with sextic dispersion relation is the same as the ghost with quartic dispersion relation:  As it is shown in \cite{arkani01,arkani02} such a coupling would have to respect the shift-symmetry of the ghost field. If for the fermionic field the symmetry $\psi\rightarrow e^{ic_{\psi}\phi/F}\psi$ is broken by a mass term $m_D\psi\bar{\psi}$, the vector coupling 
\begin{equation} \label{s04eq01}
    \frac{c_{\psi}}{F}\bar{\psi}\gamma^{\mu}\psi\partial_{\mu}\phi
\end{equation}
could be removed but the axial coupling
\begin{equation}\label{s04eq02}
    \frac{c_{\psi}}{F}\bar{\psi}\gamma^{\mu}\gamma^5\psi\partial_{\mu}\phi
\end{equation}
give rise to a different dispersion relation for the particles and anti-particles through the term, $m_\mathrm{eff}\bar{\psi}\gamma^{0}\gamma^{5}$,
\begin{equation}\label{s04eq03}
\omega=\sqrt{(|p|\pm m_\mathrm{eff})^2+m_D^2}
\end{equation}
where $m_\mathrm{eff} \equiv M^2/F$. Above, plus and negative signs are, respectively for the left helicity particle and right-helicity antiparticles. If the earth is moving with respect to the background in which the ghost field is isotropic by the velocity $\vec{v}_{\mathrm{earth}}\sim 10^{-3}$, the interaction becomes
\begin{equation}\label{s04eq04}
  m_\mathrm{eff} \bar{\psi}\Vec{\gamma}{\gamma}^5\psi\cdot\vec{v}_{\mathrm {earth}}\sim \mu \vec{S}\cdot\vec{v}_{\mathrm{earth}}\,,
\end{equation}
where there are experimental bounds on the parameter $m_\text{eff}$. For example, for electrons $m_\mathrm{eff}\lesssim10^{-25}$ GeV and for protons and neutrons $m_\mathrm{eff}\lesssim10^{-24}$ \cite{Heckel:1999sy,Phillips:2000dr,Cane:2003wp,Bluhm:2003ne}.

A new spin-dependent long-range force is expected from the exchanges of $\pi$ Goldstone boson. In the non-relativistic limit, $\pi$ has derivative coupling to the spin,
\begin{equation}\label{s04eq05}
\Delta \mathcal{L} \sim \frac{1}{F} \vec{S} \cdot \nabla \pi,
\end{equation} 
where $\vec{S}$ is the spin operator.The potential for such an interaction with a $\omega^2\propto k^6$ dispersion relation could be computed via the prescription \cite{wil}
\begin{equation}\label{s04eq06}
V = \text{coupling constant} \int \frac{\mathrm{d}^3k}{(2 \pi)^3} (\text{vertex}_1) (\text{propagator})(\text{vertex}_2).
\end{equation}
The vertex form here is $-i\vec{k}\cdot \vec{S}$ that arises from \eqref{s04eq05} Lagrangian. In the $\omega \rightarrow 0$ limit, the potential that arises from the dispersion relation \eqref{s02eq12} will be (more details in Appendix \ref{AppenA})
\begin{equation}\label{s04eq07}
\begin{split}
 V &= \frac{M^4}{\bar{S} F^2} (\vec{S}_1\cdot \nabla)(\vec{S}_2\cdot \nabla) \int \frac{\mathrm{d}^3k}{(2 \pi)^3} \frac{e^{i\vec{k} \cdot \vec{r}}}{k^6} = \frac{M^4}{32 \pi \bar{S} F^2} \left[\vec{S}_1 \cdot \vec{S}_2 + (\vec{S}_1 \cdot \hat{r})(\vec{S}_2 \cdot \hat{r})\right] r\,.
\end{split}
\end{equation}
This gives rise to a spin-dependent but distance-independent force. Above, it has been assumed that the sources are static in the time scale longer than
\begin{equation}\label{s04eq08}
    \tau\sim \omega^{-1}\sim M^2 r^3\,,
\end{equation}
where $r$ is the separation of the sources. 

\section{Coupling the Ghost Condensate to Gravity}

Although one can consider the Lagrangian for the ghost field minimally coupled to gravity, it is useful to work in the unitary gauge where the modification of the gravity is explicit. One can break the time-diffeomorphism by choosing the gauge where
\begin{equation}\label{s05eq01}
\phi(t,x)=t\,
\end{equation}
which breaks the 4d diffeomorphisms into spatial diffeomorphisms. One can write down the unitary gauge action with the terms that respect the residual symmetry,
\begin{eqnarray}\label{s05eq02}
    t'&=&t\,,\nonumber \\
    x'^{i}&=&x^i+\xi^i(t,x)\,.
\end{eqnarray}

Looking at infinitesimal deviations around the flat spacetime, $g_{\mu\nu} = \eta_{\mu\nu}+h_{\mu\nu}$, one of the invariants is $h_{00}$. One can restore the full diffeomorphisms by performing the Stueckelberg trick, i.e. by performing a broken $\xi^{0}$ and then promoting it to the field $\pi$,
\begin{equation}\label{s05eq03}
    h_{00}\rightarrow h_{00}-2\partial_{0}\pi\,, \quad  h_{0i}\rightarrow h_{0i}-\partial_{i}\pi\,,\quad h_{ij}\rightarrow h_{ij}
    \end{equation}
As pointed out in \cite{arkani01}, the kinetic term proportional to 
\begin{equation}\label{s05eq04}
    \frac{1}{8}\int  M^4 h_{00}^2
\end{equation}
generates  the kinetic term for $\pi$, i.e. $\dot{\pi}^2$. The extrinsic curvature terms proportional to $K_{ii}^2$ and $K_{ij}$ generate the $(\nabla^2 \pi)^2$, since
\begin{equation}\label{s05eq05}
     K_{ij}\rightarrow K_{ij}+\partial_{i}\partial_{j}\pi
\end{equation}
So a term proportional to $K_{ii}^2$ and $K_{ij}^2$ would induce the quartic dispersion relation. We adjust the coefficients of these terms to be subdominant with respect to the sixth-order terms in the dispersion relation. Noting how $K_{ij}$ transforms, Eq. \eqref{s05eq05}, we need terms like $\nabla_i K_{jk} \nabla_i K_{jk}$, or more exactly the combination
\begin{equation}\label{s05eq06}
  S=-\int \mathrm{d}^3 x \mathrm{d}t \left(\frac{\sigma_1}{2}\nabla_i K_{jk} \nabla_i K_{jk}+\frac{\sigma_2}{2}(\nabla_i K_{jj})^2+\frac{\sigma_3}{2}\nabla_{i}K_{ij}\nabla_{l}K_{lj}+\frac{\sigma_4}{2}\nabla_{i}K_{ij}\nabla_{j}K_{ll} \right)
\end{equation}
Reintroducing $\pi$ in the above Lagrangian one obtains
\begin{equation}\label{s05eq07}
S=\int -\frac{\sigma}{2}\left[ (\nabla^3 \pi)^2 + \cdots \right]
\end{equation}
 where $\sigma \equiv \sum_{i=1}^4\sigma_i$. Following \cite{arkani01} In a de Sitter background the effective action in the unitary gauge will become
  \begin{eqnarray}\label{s05eq08}
  S&=&\int \mathrm{d}^3 x \mathrm{d}t \sqrt{\gamma}\left(\frac{M^4}{8}(X-1)^2-\frac{\sigma_1}{2}\gamma^{il}\nabla_i K_{jk} \nabla_l K^{jk}-\frac{\sigma_2}{2}\gamma^{il}\nabla_i K^{j}_{\ j}\nabla_l K^{j}_{\ j}-\frac{\sigma_3}{2}\nabla_{i}K_{ij}\nabla_{l}K_{lj}\right.\nonumber\\&-&\left.\frac{\sigma_4}{2}\nabla_{i}K_{ij}\nabla_{j}K_{ll}\right).
  \end{eqnarray}
It should be noted here the perpendicular unit vector $n_\mu$, the induced metric $\gamma_{\mu \nu}$ and $K_{\mu \nu}$ are defined as follows,
\begin{align}
n_\mu & = \frac{\partial_{\mu} \phi}{\sqrt{X}} \rightarrow \frac{\delta_{\mu}^{0}}{\sqrt{g^{00}}}, \label{s05eq09}  \\
 \gamma_{\mu \nu} & = g_{\mu \nu} - n_\mu n_\nu \label{s05eq10},
\\K_{\mu \nu} & = \gamma^\rho_{\ \mu} \nabla_\rho n_\nu. \label{s05eq11}
\end{align}
The term proportional to $(X-1)^2$ generates the kinetic term, $\dot{\pi}^2$ for the ghost field.

Let us now look at the modification of gravity in the ghost condensate at the linearized level. Using the perturbed metric in the longitudinal gauge
\begin{equation}\label{s05eq12}
\mathrm{d}s^2 = -(1+ 2\Phi(t,x))\mathrm{d}t^2 +(1-2 \Psi(t,x))\delta_{i j} \mathrm{d}x^i \mathrm{d}x^j\,,
\end{equation}
the Einstein-Hilbert action
\begin{equation}\label{s05eq13}
S_{\mathrm{EH}}=\frac{M^2_{\mathrm{Pl}}}{2}\int \mathrm{d}^4x\sqrt{-g} R
\end{equation}
up to second order in $\Phi$ and $\Psi$ and after Fourier transform takes the form
\begin{equation} \label{s05eq14}
    \mathcal{L}_\text{EH}  =M_{\text{Pl}}^2\left[-3 \dot{\Psi}^2-2 k^2 \Psi \Phi + k^2\Psi^2\right]\,.
\end{equation}
In the Newtonian limit $\omega^2\ll k^2$, one obtains $\Phi=\Psi$. Including the ghost condensate, the Lagrangian at the quadratic level assuming Newtonian limit $ \omega^2\ll k^2$ is
\begin{equation}\label{s05eq15}
\mathcal{L}_\text{eff} =\frac{1}{2} M^4 \left(\Phi - \dot{\pi} \right)^2 -\frac{\sigma}{2}\left( k^6 \pi^2 + . . . \right).
\end{equation}
Introducing the canonical variables, $M^2 \pi \rightarrow \pi_c, \sqrt{2} M_\text{Pl} \Phi \rightarrow \Phi_c$ and setting $\mu \equiv M^2/\sqrt{2}M_\text{Pl}$, the kinetic matrix of the Lagrangian becomes
\begin{equation}\label{s05eq16}
\frac{1}{2}\left(
\begin{array}{cc}
-k^2 + \mu^2 & i \mu \omega \\
- i \mu \omega  & \omega ^2 -\frac{\sigma}{M^4} k^6 \\
\end{array}
\right).
\end{equation}
The determinant of \eqref{s05eq16} gives the dispersion relation
\begin{equation}\label{s05eq17}
\omega ^2 = \frac{\sigma}{M^4}\left(k^6 - \mu^2 k^4 \right).
\end{equation}
Here, the $k^4$ term arises due to the mixing with gravity.  Contrary to the ghost condensate with quartic dispersion relation, mixing with gravity adds a negative quartic term to the dispersion relation.
$\omega^2$ takes negative values for momenta $k < \mu $, which means we face an instability similar to the Jeans instability. Like the ghost with quartic dispersion relation, it is expected that this instability is removed by Hubble friction in an expanding background.

The width of the instability in $\omega$ gives a timescale, $t_c$, that a change in the gravitational potential happens at length scales, $\sim \mu^{-1}$. For this purpose, we should find the largest imaginary magnitude of $\omega$. Minimizing the equation \eqref{s05eq17} gives the largest imaginary magnitude of $\omega$,
\begin{equation}\label{s05eq19}
\omega_\text{ins} = i \frac{1}{3}\sqrt{\frac{\sigma}{6}} \frac{M^4}{M_\text{Pl}^3} = i\frac{2}{3}\sqrt{\frac{\sigma}{6}} \frac{\mu^2}{M_\text{Pl}}  \equiv i \Upsilon.
\end{equation}
In order to obtain the modified gravity potential, we need to obtain the $\langle \Phi_c\Phi_c\rangle$ operator, which can be obtained by inverting the kinetic matrix \eqref{s05eq16}. One obtains
\begin{equation}\label{s05eq21}
\frac{1}{k^2}\left(-1+\frac{\frac{\sigma \mu^2}{M^4}  k^4}{-\frac{\sigma}{M^4} k^6+\frac{\sigma \mu^2}{M^4} k^4 + \omega ^2}\right).
\end{equation}
\begin{figure}[t!] 
\centering
\includegraphics[width=.999\textwidth]{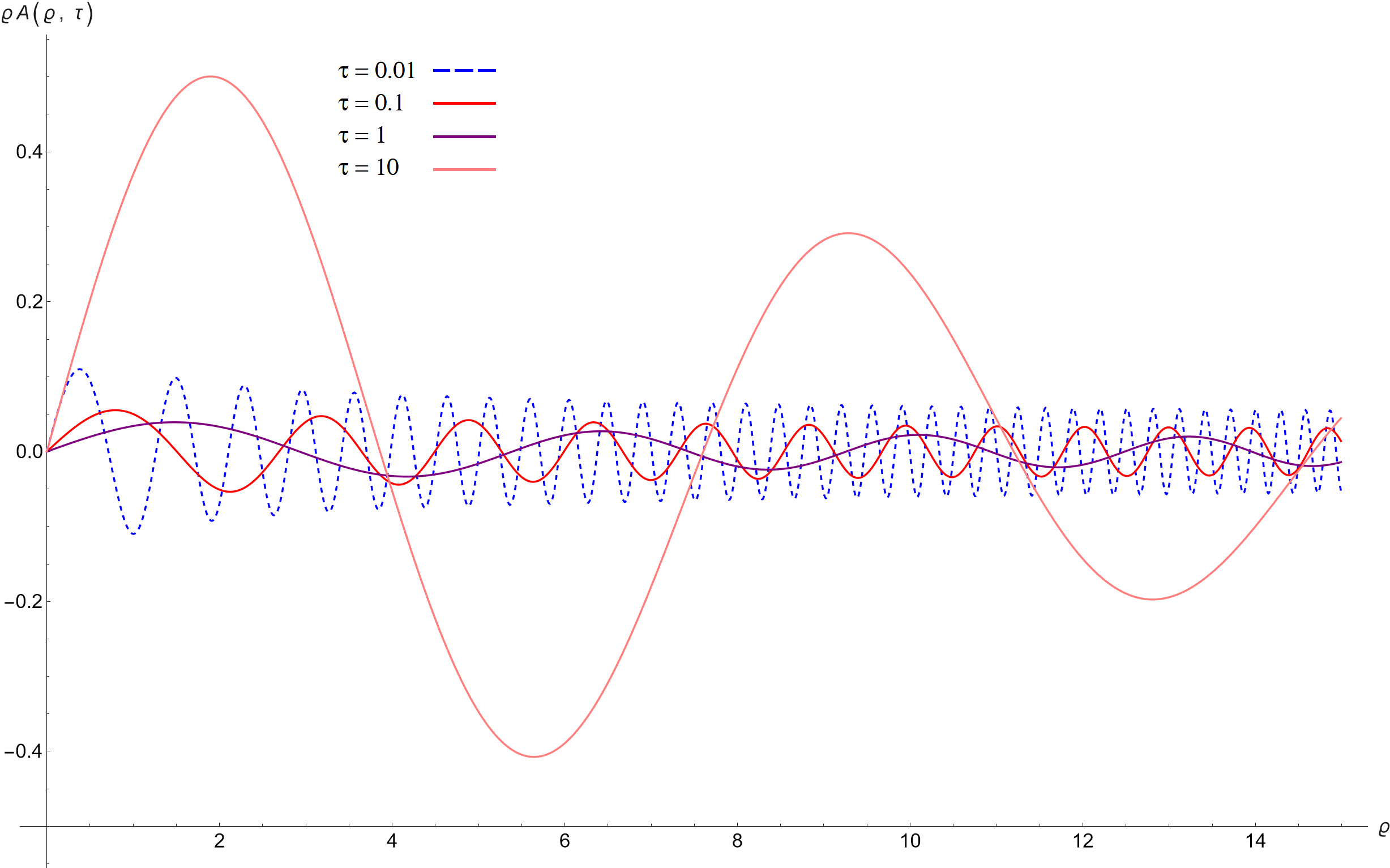}
\caption{The modification part of the gravitational potential is plotted as $\varrho A(\varrho)$. Axes are rescaled. Blue dashed line: $\uptau = 10^{-2} \Upsilon$. Red line: $\uptau = 10^{-1} \Upsilon$. Purple line: $\uptau = \Upsilon$. Pink line: $\uptau = 10^{1} \Upsilon$.}
\label{fig01}
\end{figure}
The factor in the parenthesis corresponds to the change in the Newtonian potential. Demanding that this change is $\mathcal{O}(1)$ one obtains
\begin{equation}
\omega^2-\frac{\sigma}{M^4}k^6\lesssim \frac{\sigma \mu^2}{M^4}k^4\,,
\end{equation}
or alternatively
\begin{equation}\label{s05eq20}
\left|\omega - \frac{\sqrt{\sigma}}{M^2} k^3\right| \lesssim \sqrt{\sigma}\frac{\mu^2}{M_\text{Pl}} = \frac{3\sqrt{6}}{2} \Upsilon
\end{equation}
where above we have assumed $k\lesssim \mu$. This suggests that there is an instability timescale, $t_c$, which can be characterized as the inverse of $\Upsilon$. There is also a characteristic length scale, $\mu^{-1}$. 
\begin{equation}
    r_c\simeq \frac{M_{\mathrm{Pl}}}{M^2}\,,\qquad t_c\sim\frac{M_{\mathrm{Pl}}^3}{\sqrt{\sigma}M^4}
\end{equation}
The time scale in the case of sixth order dispersion relation is about a factor $M_{\mathrm{Pl}}/M$ larger than the time scales in the quartic dispersion relation. Like the quartic dispersion relation, due to the slow-down of the signal at large length scales, $r_c\ll t_c$.

Now, let us consider a test mass point-like particle that appears at the origin at time $t =0$. It means that the density of the source is $\rho(r,t) = \delta(\vec{r}) \theta(t)$. The inverse Fourier transform of the $\Phi$ propagator \eqref{s05eq21} gives the potential of gravity in the position space. For the Fourier integration, we define the following dimensionless variables,
\begin{equation}\label{s05eq22}
\begin{split}
\kappa \equiv \mu^{-1} k, \quad \Omega \equiv \frac{2}{3\sqrt{6}} \frac{\omega}{\Upsilon} , \quad  \varrho \equiv \mu r,  \quad  \uptau \equiv \frac{3\sqrt{6}}{2} \Upsilon t.
\end{split}
\end{equation}

The result is made of two parts: the first one give rises to the standard gravity, $\Phi_\text{N}$, whereas the second part arises from modifications to gravity due to the ghost condensate, $\Phi_\text{Mod}$, which means
\begin{align}
&\Phi  = \Phi_\text{N} + \Phi_\text{Mod}, \label{s05eq23} \\
& \Phi_\text{N} = -\frac{M_\text{Pl}^{-2}}{r}, \quad  \Phi_\text{Mod}= M_\text{Pl}^{-2} A(r, t), \label{s05eq24}
\end{align}
where
\begin{equation}\label{s05eq25}
\begin{split}
A(r, t) & =\int \frac{\mathrm{d}^3 \kappa \ \mathrm{d} \Omega}{(2 \pi)^4} \frac{e^{i \Omega \uptau} e^{i \vec{\kappa} \cdot \vec{\varrho}}}{\kappa^2}\frac{\kappa^4}{\kappa^4 -\kappa^6+ \Omega^2} \\ 
& = \frac{1}{2 \pi^2 \varrho} \left( \int^1_0 \mathrm{d} \kappa \frac{\kappa \sin \left(\kappa \varrho\right) \sinh \left(\uptau \kappa^2 \sqrt{1-\kappa^2}\right)}{\sqrt{1-\kappa^2}}  + \int^{\infty}_{1} \mathrm{d} \kappa  \frac{\kappa \sin \left(\kappa \varrho\right) \sin \left(\uptau \kappa^2 \sqrt{\kappa^2 - 1}\right)}{\sqrt{\kappa^2- 1}} \right).
\end{split}
\end{equation}
The result of integral $A(r,t)$ is not analytical, and we have to calculate it numerically. The result is shown in Fig. \ref{fig01}. The numerics suggest that the modification will be significant only for $\uptau \gtrsim 1$, or in other words, $t \gtrsim t_c$. For larger values of $\uptau$ the modulated oscillations have a larger periodicity and amplitude. The result is  indicative of the slow-varying amplitude of $r A(r,t)$ and so $\Phi_\text{Mod}$. The effect of the modification can be observed at longer distances; however, the larger value of this modification takes place at $\varrho \gtrsim 1$ ($r \gtrsim r_c$). With increasing $\uptau$, the period of oscillations of $r A(r,t)$ increases. Only for $\uptau > 1$, we will have substantial modification to the Newtonian potential. 

\section{Discussion and Further Work}
Ghost condensate can be regarded as the origin of the universe current \cite{arkani01} and primordial accelerations \cite{Arkani-Hamed:2003juy}. The original work argues that the dispersion relation is at most $\omega^2 \sim k^4$, and higher order terms like $k^6$ is not viable because of the strong-coupling of the non-interacting terms in the IR regime. This argument was challenged in \cite{ashoorioon18} during inflation as the dispersion relation could be supplemented with lower powers of the dispersion relation as the mode stretches due to the cosmic expansion. Here we argue for the pure $k^6$ dispersion relation for the ghost dark energy, as the primordial acceleration is a relatively recent phenomenon. Thus the IR modes have not really stretched out to infinitely large wavelengths. One can quantify this by computing the amplitude of the three-point function and demanding it to be sub-dominant with respect to the two-point function. It can be shown that for our universe with the history of acceleration domination around $z_c\sim 0.3$, this will lead to a very mild bound on the coefficient of the sixth-order dispersion relation. We showed that the strong coupling only becomes important when the redshift of the acceleration of the universe would become $\sim 10^{33}$.

We also focused on the possibility of direct interaction between the ghost condensate with the sextic dispersion relation and the standard model particles. As expected, we obtained a spin-dependent Lorentz-violating force that has no dependency on distance. This result is fascinating because a distance-independent force can become dominant when all other forces and interactions weaken. Accordingly, we can even imagine a constant (non-gravitational) force in all spacetime that competes with gravity on large scales and makes some interesting phenomena. At the same time, we know that the ghost condensate modifies gravity separately.

In this work, the direct mixing between the ghost sector and gravity in the IR shows that the Newtonian potential receives modifications with oscillatory behaviors at large distances and especially at late times, similar to what the standard ghost with quartic dispersion exhibits in \cite{arkani01}. Compared to the ghost condensate theory with the quartic dispersion relation, here, the fall-off of the amplitude of oscillations happens at much later times.

With the assumption of the maximally quartic dispersion relation, effective field theory of dark energy has been founded \cite{Gubitosi:2012hu}. It would be interesting to investigate the ghost dark energy with the sixth-order dispersion relation in de Sitter space. Also, it is interesting to assess the possibility of the ghost field with the sextic dispersion relation as dark matter \cite{Furukawa:2010gr}. The phenomenology of the  ghost sector near a black hole \cite{Frolov:2004vm, Krotov:2004if, Mukohyama:2005rw} is another avenue to pursue. 

\acknowledgments
We would like to thank S. Mukohyama, M. B. Jahani Poshteh, A. Rostami and M. M. Sheikh-Jabbari for the helpful discussions. This project has received funding /support from the European Union’s Horizon 2020 research and innovation programme under the Marie Skłodowska -Curie grant agreement No 860881-HIDDeN.

\appendix
\section{Appendix }\label{AppenA}
We would like to compute 
\begin{equation} \label{eq-ap01}
 (\vec{S}_1\cdot \nabla)(\vec{S}_2\cdot \nabla) f(r),
\end{equation}
where 
\begin{equation}\label{eq-ap02}
f(r) =\int \frac{\mathrm{d}^3k}{(2 \pi)^3} \frac{e^{i\vec{k}\cdot \vec{r}}}{k^n}\,.
\end{equation}
The following two vector identities will be used 
\begin{align}
\nabla(\vec{A} \cdot \vec{B}) & = (\vec{A} \cdot \nabla)\vec{B} + (\vec{B} \cdot \nabla)\vec{A} + \vec{A} \times (\nabla \times \vec{B})+ \vec{B} \times (\nabla \times \vec{A}) \label{eq-ap03}
\\
(\vec{a} \cdot \nabla) [\hat{r} f(r)] & = \frac{f(r) [ \vec{a} - \hat{r} (\vec{a} \cdot \hat{r}) ]}{r}  + \hat{r}(\vec{a} \cdot \hat{r}) \partial_r f(r) \label{eq-ap04}
\end{align}
where $\vec{a}$ is an arbitrary vector that is not dependent on $r$. $\vec{A}$ and $\vec{B}$ are two arbitrary vector functions \cite{arfken}.
In the first step, we note that
\begin{equation}
\label{eq-ap05}
(\vec{S}_1\cdot \nabla)(\vec{S}_2\cdot \nabla) f(r) = \vec{S}_1 \cdot \left[ \nabla [ \vec{S}_2\cdot (\nabla f(r))] \right] 
\end{equation}
Using \eqref{eq-ap03}, we obtain
\begin{equation}
\label{eq-ap06}
\begin{split}
(\vec{S}_1\cdot \nabla)(\vec{S}_2\cdot \nabla) f(r) = \vec{S}_1 \cdot & \left[ (S_2 \cdot \nabla) \nabla f(r) + (\nabla f(r) \cdot \nabla) \vec{S}_2 + \vec{S}_2 \times (\nabla \times \nabla f(r)) \right. \\ & \left. + \nabla f(r) \times (\nabla \times \vec{S}_2) \right]\,.
\end{split}
\end{equation}
Since $\vec{S}_{1,2}$ are not dependent on $r$ and $\nabla \times \nabla f(r) = 0$, only the first term is non-zero in the RHS of equation \eqref{eq-ap06}. Also as $f$ is a function of $|\vec{r}|$, $\nabla f(r)$ reduces to $\hat{r} \partial_r f(r)$ and we can now utilize \eqref{eq-ap03} identity. Hence
\begin{equation}\label{eq-ap07}
\begin{split}
(\vec{S}_1\cdot \nabla)(\vec{S}_2\cdot \nabla) f(r) & =  \vec{S}_1 \cdot \left[(S_2 \cdot \nabla) \hat{r} \partial_r f(r) \right] \\ & = \vec{S}_1 \cdot \left[ \frac{\partial_r f(r)}{r} \left( \vec{S}_2 - (\vec{S}_2 \cdot \hat{r}) \hat{r} \right) + (\vec{S}_2 \cdot \hat{r}) \hat{r} \partial^2_r f(r)  \right]  \\ & = \frac{(\vec{S}_1 \cdot \hat{r})(\vec{S}_2 \cdot \hat{r})}{r}[r \partial_r^2 - \partial_r]f(r) + \frac{\vec{S}_1 \cdot \vec{S}_2}{r}\partial_r f(r).
\end{split}
\end{equation}

Now we compute $f(r)=\int \frac{\mathrm{d}^3k}{(2 \pi)^3} \frac{e^{i\vec{k}\cdot \vec{r}}}{k^n}$.
In the spherical coordinate,
\begin{equation}\label{eq-ap08}
\int \frac{\mathrm{d}^3k}{(2 \pi)^3} \frac{e^{i\vec{k}\cdot \vec{r}}}{k^n} = \frac{1}{(2\pi)^3} \int^{\infty}_0 \int^{1}_{-1} \int^{2\pi}_0 \mathrm{d} k  \ \mathrm{d}(\cos (\theta)) \ \mathrm{d} \phi \ k^2 \frac{e^{ikr cos(\theta)}}{k^n}.
\end{equation} 
One first integrate over $\phi$ and $\cos(\theta)$,
\begin{equation}\label{eq-ap09}
\int \frac{\mathrm{d}^3k}{(2 \pi)^3} \frac{e^{i\vec{k}\cdot \vec{r}}}{k^n} = \frac{1}{(2\pi)^3}\frac{2 \pi}{i r} \int_{0}^{\infty} \mathrm{d} k \, \frac{k^2}{k^n} \left( \frac{e^{ikr}}{k} + \frac{e^{-ikr}}{-k} \right)\,.
\end{equation}
Using the change of parameter $k \rightarrow -k$ in the second half of the integrand, 
\begin{equation}\label{eq-ap10}
\int \frac{\mathrm{d}^3k}{(2 \pi)^3} \frac{e^{i\vec{k}\cdot \vec{r}}}{k^n} = \frac{1}{(2\pi)
^2}\frac{1}{i r} \int_{-\infty}^{\infty} \mathrm{d} k \, \frac{e^{ikr}}{k^{n-1}}.
\end{equation}
Afterward, the Cauchy integral theorem
\begin{equation}\label{eq-ap11}
\frac{\partial^m g(z)}{\partial z^m}\Big|_{z = u} = \frac{m!}{2 \pi i} \oint \mathrm{d}z \frac{g(z)}{(z-u)^{m+1}},
\end{equation}
can be used, where $u$ is a simple pole \cite{arfken}. Using this theorem, the result of the integration \eqref{eq-ap09} will become
\begin{equation}\label{eq-ap12}
\begin{split}
\int \frac{\mathrm{d}^3k}{(2 \pi)^3} \frac{e^{i\vec{k}\cdot \vec{r}}}{k^n}  =  \frac{1}{(2\pi)^2}\frac{1}{i r} \int_{-\infty}^{\infty} \mathrm{d} k \, \frac{e^{ikr}}{k^{n-1}}=\frac{1}{(n-2)!}\frac{1}{4\pi r} \frac{\partial^{n-2}}{\partial k^{n-2}} e^{ikr}\Big|_{k = 0}.
\end{split}
\end{equation}

\end{document}